\documentstyle{mn}

\newif\ifAMStwofonts

\ifoldfss
  \ifCUPmtlplainloaded \else
    \NewTextAlphabet{textbfit} {cmbxti10} {}
    \NewTextAlphabet{textbfss} {cmssbx10} {}
    \NewMathAlphabet{mathbfit} {cmbxti10} {} 
    \NewMathAlphabet{mathbfss} {cmssbx10} {} 
  \fi
  \ifAMStwofonts
    \ifCUPmtlplainloaded \else
      \NewSymbolFont{upmath} {eurm10}
      \NewSymbolFont{AMSa} {msam10}
      \NewMathSymbol{\upi}     {0}{upmath}{19}
      \NewMathSymbol{\umu}     {0}{upmath}{16}
      \NewMathSymbol{\upartial}{0}{upmath}{40}
      \NewMathSymbol{\leqslant}{3}{AMSa}{36}
      \NewMathSymbol{\geqslant}{3}{AMSa}{3E}

      \let\leq=\leqslant 
      \let\geq=\geqslant 
    \fi
  \fi
\fi 

\ifnfssone
  \newmathalphabet{\mathit}
  \addtoversion{normal}{\mathit}{cmr}{m}{it}
  \addtoversion{bold}{\mathit}{cmr}{bx}{it}
  \newmathalphabet{\mathbfit} 
  \addtoversion{normal}{\mathbfit}{cmr}{bx}{it}
  \addtoversion{bold}{\mathbfit}{cmr}{bx}{it}
  \newmathalphabet{\mathbfss} 
  \addtoversion{normal}{\mathbfss}{cmss}{bx}{n}
  \addtoversion{bold}{\mathbfss}{cmss}{bx}{n}
  \ifAMStwofonts
    \ifCUPmtlplainloaded \else
      \UseAMStwoboldmath
      \makeatletter
      \new@mathgroup\upmath@group
      \define@mathgroup\mv@normal\upmath@group{eur}{m}{n}
      \define@mathgroup\mv@bold\upmath@group{eur}{b}{n}
      \edef\UPM{\hexnumber\upmath@group}
      \new@mathgroup\amsa@group
      \define@mathgroup\mv@normal\amsa@group{msa}{m}{n}
      \define@mathgroup\mv@bold\amsa@group{msa}{m}{n}
      \edef\AMSa{\hexnumber\amsa@group}
      \makeatother
      \mathchardef\upi="0\UPM19
      \mathchardef\umu="0\UPM16
      \mathchardef\upartial="0\UPM40
      \mathchardef\leqslant="3\AMSa36
      \mathchardef\geqslant="3\AMSa3E

      \let\leq=\leqslant 
      \let\geq=\geqslant 
    \fi
  \fi
\fi 

\ifnfsstwo
  \DeclareMathAlphabet{\mathbfit}{OT1}{cmr}{bx}{it}
  \SetMathAlphabet\mathbfit{bold}{OT1}{cmr}{bx}{it}
  \DeclareMathAlphabet{\mathbfss}{OT1}{cmss}{bx}{n}
  \SetMathAlphabet\mathbfss{bold}{OT1}{cmss}{bx}{n}
  \ifAMStwofonts
    \ifCUPmtlplainloaded \else
      \DeclareSymbolFont{UPM}{U}{eur}{m}{n}
      \SetSymbolFont{UPM}{bold}{U}{eur}{b}{n}
      \DeclareSymbolFont{AMSa}{U}{msa}{m}{n}
      \DeclareMathSymbol{\upi}{0}{UPM}{"19}
      \DeclareMathSymbol{\umu}{0}{UPM}{"16}
      \DeclareMathSymbol{\upartial}{0}{UPM}{"40}
      \DeclareMathSymbol{\leqslant}{3}{AMSa}{"36}
      \DeclareMathSymbol{\geqslant}{3}{AMSa}{"3E}

      \let\leq=\leqslant 
      \let\geq=\geqslant 
    \fi
  \fi
\fi 

\ifCUPmtlplainloaded \else
  \ifAMStwofonts \else 
    \def\upi{\pi}
    \def\umu{\mu}
    \def\upartial{\partial}
  \fi
\fi

\title{EVN observations of low-luminosity flat-spectrum AGNs}
\author[A. Caccianiga et al.]
       {A. Caccianiga,$^1$ \thanks{caccia@brera.mi.astro.it}
        M.J.M. March\~a,$^1$ A. Thean,$^2$ 
	J. Dennett-Thorpe$^3$\\
        $^1$CAAUL, Observat\'orio Astron\'omico de Lisboa, Tapada da Ajuda, 
	1349-018 Lisboa, Portugal \\
	$^2$ Istituto di Radioastronomia del CNR, Via P. Gobetti 101, 
	I-40129 Bologna, Italy \\
	$^3$ University of Groningen, Kapteyn Institute, 9700AV Groningen, 
	The Netherlands\\
        }
\date{}

\pagerange{\pageref{firstpage}--\pageref{lastpage}}
\pubyear{1994}

\begin{document}

\input{psfig.txt}
\maketitle

\label{firstpage}

\begin{abstract}
We present and discuss the results of VLBI (EVN) observations 
of three low-luminosity (P$_{5~GHz}<$10$^{25}$ W Hz$^{-1}$)
Broad Emission Line AGNs carefully selected from a sample of flat spectrum
radio sources (CLASS). Based on the total and the extended radio power 
at 5~GHz and at 1.4~GHz respectively, these objects should be technically 
classified as radio-quiet AGN and thus the origin of their
radio emission is not clearly understood. The VLBI observations
presented in this paper 
have revealed compact radio cores which imply a lower limit on the 
brightness temperature of about 3$\times$10$^8$~K. This result
rules out a thermal origin for the radio emission and strongly 
suggests an emission mechanism similar to that observed in more 
powerful radio-loud AGNs. Since, by definition, the three objects
show a flat (or inverted) radio spectrum between 1.4~GHz and 8.4~GHz, 
the observed radio emission could be relativistically beamed. 
Multi-epoch VLBI observations can confirm this possibility 
in two years time.

\end{abstract}


\section{introduction}
The origin of the bimodality in radio luminosities of the
quasar population remains a puzzle in AGN studies. This
is due to the fact that, apart from the strength of the 
radio emission, radio-quiet and radio-loud  (RQ and RL, respectively) 
AGNs share many of the same properties across the electromagnetic spectrum. 
It has been
proposed (Terlevich et al. 1992) that in the case of the RQ AGNs, the radio 
emitting
mechanism could be due to starburst phenomena in a very dense environment,
whereas in the case of RL AGN the radio emission is due to the presence of
radio jets. However, the detection of strong and compact (at VLBI resolution) 
nuclei in many  RQ AGNs (Blundell \& Beasley 1998) suggests that, at least in 
some RQ AGN, a non-thermal jet-like engine is at work. 

With the increasing sensitivity of radio surveys,
it has been shown that some local RQ AGNs (Seyfert galaxies) actually 
possess weaker
versions of the jets seen in their radio-loud counterparts (e.g 
Mazzarella et al. 1991; Pedlar et al. 1993; Kukula et al. 1993; 
Kukula et al. 1996; Falcke, Wilson \& Simpson 1998; Capetti et al. 1999; 
Thean et al. 2000). 
The study of the properties of these weak jets is clearly of
fundamental importance to understand why in some
objects the formation and/or the evolution of large-scale jets 
is inhibited. In particular, it would be interesting to know if the
weak radio jets found in RQ objects are relativistic, like
those found in RL AGNs. If so, when the source is seen along the jet, 
we expect to see some relativistic effects, like superluminal (SL) motion
or high brightness temperature ($>$10$^{11}$ K) radio components, 
similar to those observed in BL Lac objects.

Falcke et al. (1996) have identified a number of AGNs, selected 
from the Palomar-Green catalog (PG)  whose radio
luminosity, if compared with the optical one, is intermediate between
the typical RQ and RL AGNs. The distinction between RQ and RL
object is commonly based on the radio-to-optical
flux ratio ({\it R}) defined as (Kellerman et al. 1994):

\begin{center}
{\it R} = S$_{5~GHz}$ / S$_{4400\AA}$ 
\end{center}

where S$_{5~GHz}$ and S$_{4400\AA}$ are respectively the radio (at 5~GHz) 
and optical (at 4400~\AA) flux densities.
RQ and RL AGNs are respectively defined 
(Kellerman et al. 1994) with {\it R} below and above 10,  
while flat radio spectrum  AGNs with 10$< R <$ 250 are classified by 
Falcke et al. (1996) as Radio Intermediate (RI) AGNs. 
Falcke et al. (1996) present 
evidence  that RI AGNs are RQ AGNs whose weak radio jet is pointing
toward the observer and, as a consequence, the radio emission
is beamed in that direction. In one of the objects selected 
by Falcke et al. (1996), namely the Seyfert galaxy III Zw 2,
evidence for a superluminal jet has been presented (Brunthaler et al. 2000).

The confirmation that RQ AGNs have relativistic
jets would be of great interest for our understanding of the
radio-loud/radio-quiet dichotomy. More than finding a few
examples of relativistic jets in RQ AGNs it would be
important to quantify the percentage of RQ AGNs that 
present this characteristic. To this end, a systematic 
search for relativistic effects in a well defined and complete 
sample of RQ/RI AGNs would be 
very useful. 

In this paper we present VLBI observations with the European 
VLBI Network (EVN) of three low redshift (z$<$0.1) RQ/RI AGNs 
carefully selected from a deep survey of flat spectrum radio sources.
The goal of this small pilot sample is to test the efficiency 
with which radio criteria may be used to find beamed RQ AGNs. 

Throughout this paper we use $H_0$=50 km s$^{-1}$ Mpc$^{-1}$ and 
q$_0$=0.

\section{Selection strategy}

In the case of radio-loud AGNs 
the ``classical'' starting point to look for beamed objects
(called ``blazars'') is a radio survey of flat-spectrum
sources, since a flat or inverted spectrum in the radio band is 
the most easily recognised indication of relativistically 
beamed emission. Until recently the radio surveys used to 
select flat-spectrum AGN have had flux limits of around 
1~Jy or higher (e.g. K\"uhr et al. 1981). Because the intrinsic
power of a RQ AGN is, on average, two or three orders of magnitude
lower than that of a RL AGN, these surveys would be unlikely
to contain a useful fraction of beamed RQ objects.

For this reason, 
we decided to look for beamed RQ/RI objects using 
the deepest available sample of flat spectrum radio sources, 
namely  the CLASS survey (Myers et al. 2001). This survey is defined 
as follows:

\begin{itemize}

\item $S_{5 GHz} \geq$ 30 mJy; 

\item $\alpha_{5}^{1.4} \leq$ 0.5 (S$\propto\nu^{-\alpha}$); 

\item 0$^{\circ}< \delta<$75$^{\circ}$

\end{itemize}

\noindent

Since a RQ/RI AGN is expected to have, on average,  
bright optical  nuclear emission, we restrict
the search to the radio sources in CLASS  with a bright optical counterpart.
A complete subsample of optically-bright objects (R$<$17.5) has been
already selected from CLASS and presented in March\~a et al. (2001). 
About 70\% of the 325 objects contained in this sample have 
a spectroscopical classification (Caccianiga et al. 2001). 
Since this sample has been designed
to find low-luminosity blazars we expect that  beamed RQ AGNs, if they
really exist, are to be found among these sources. 

\begin{figure}
\psfig{figure=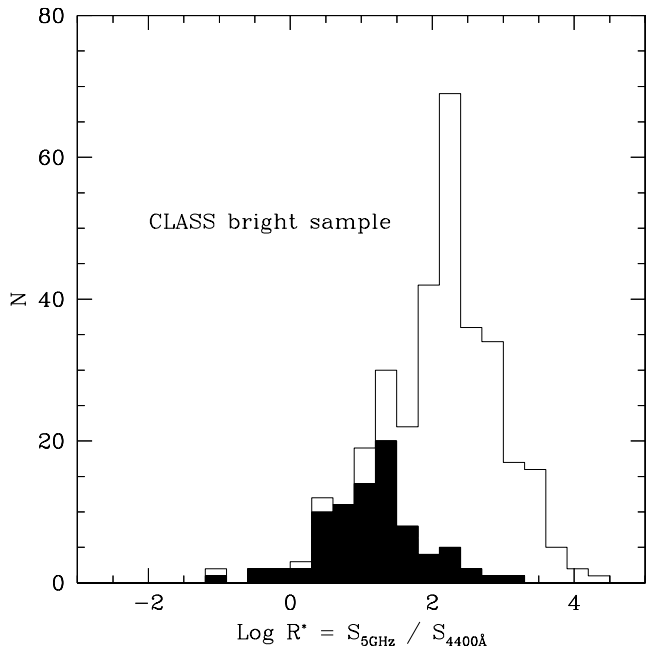,height=9cm,width=9cm}
\caption{Radio-to-optical flux ratio (R$^*$) for the 
325 objects in the CLASS bright sample computed with
the observed optical magnitude (galaxy+nucleus).
The shaded area
includes only the low redshift (z$<$0.1) sources.}
\label{r_param}
\end{figure}

For the 325 objects we have computed a ``total'' R$^*$ parameter 
(see Figure~\ref{r_param}) defined
on the basis of the observed power at 5~GHz and of the observed
red magnitude (m$_R$). Since the m$_R$ includes both the galaxy and 
the nuclear emission, the R$^*$  parameter differs from the
nuclear R. For a moderate contribution from 
the host galaxy the value of R$^*$ tends to be equal to the nuclear R 
but, for a significant amount of starlight, the value of R$^*$ can
be considerably lower than R. 

Thus, the objects in Figure~\ref{r_param} with R$^*$ below 
250 may well contain 
objects that would be classified as RL if optical AGN
contribution alone were used to calculate R.
At present, the computation of the nuclear R parameter 
for the sources in our sample is not possible since it requires
additional high  resolution data (e.g. HST) to estimate the 
optical nuclear contribution. However, since the value of 
R$^*$ is an upper limit on R, we can confidently say that 
all the RQ/RI objects present in the CLASS survey must be found among 
the objects with  R$^*\leq$250, although they are probably mixed up 
with some RL AGNs.

In order to exclude from the sample the RL AGNs we have 
used an additional selection criterion based on the radio power. 
Since the radio power at 5~GHz is expected
to be affected by beaming, we have decided to use the extended 
radio power at 1.4~GHz. About 40\% of the sources fall in the region of
sky covered by the FIRST survey (Becker et al. 1995) which is made with 
VLA in B-configuration. The relatively small beam size
(FWHM of 5$^{\prime\prime}$ which corresponds to linear sizes below 
10 kpc, for z$<0.1$)
gives the possibility of measuring the extended radio flux at 1.4~GHz:

\begin{center}
S$^{ext}$ = S$^{int}$ -- S$^{pk}$
\end{center} 

where S$^{int}$ and S$^{pk}$ are the integrated and the peak fluxes given
in the FIRST catalogue. 

Thus, among the   sources contained in the CLASS bright sample and falling 
in the region of sky covered by FIRST, we have selected our best 
candidates with the following additional criteria:

\begin{enumerate}

\item low-redshift objects (z$\leq$0.1);

\item Extended radio power at 1.4~GHz $< 5\times$10$^{22}$ W/Hz

\end{enumerate}

The reason why we have selected low-z objects is that
they allow a more detailed analysis and, for instance, any 
superluminal motion can be detected after a few years.

In total, 22 objects fulfill these selection criteria. 
We consider this list of 22 objects the  {\it sample
of beamed RQ AGN candidates}. All these objects have R$^*<$250
and all but one have R$^*<$100. Moreover, the observed
radio power at 5~GHz for all these objects is below 
10$^{25}$ W Hz$^{-1}$ which is the value often 
used to distinguish between RL and RQ AGNs (Kellermann et al. 
1994). If the observed power is relativistically beamed,
the intrinsic power is expected to be even lower than
the measured value.

\section{Target selection for the EVN observations}

The aim of this work is to study the properties of the selected 
objects 
looking for evidence of relativistic effects. The first
step is a radio observation at VLBI resolution, in order to
confirm (or rule out) a non-thermal origin for the observed
radio emission. In general, a simple detection of the majority
of the flux observed in the VLA map at VLBI
resolution (beamsize of few m.a.s) would imply a high
brightness temperature ruling out the hypothesis of a thermal 
emission. The second step is the collection of more
epochs of VLBI observation in order to look for superluminal motions.

As a pilot study, we have selected 3 objects from the 22 candidates,
and observed these with the EVN. 
Since, according to the Unified Schemes (e.g. Wills 1999),
a beamed source is
supposed to be oriented in such a way that we can see the Broad Line
Region, we expect that a good candidate for beaming would have 
broad emission lines in the optical spectrum.
Therefore, among the 22 objects in the sample,  
we have selected 3 sources that show  broad emission lines. 

The list of the observed targets is presented in Table~1
along with the radio position measured with the VLA (A-array) 
at 8.4~GHz (column~2), NVSS flux density at 1.4~GHz (column~3), 
GB6 flux density at 5~GHz (Column~4), VLA flux density at 8.4~GHz 
(column~5), redshift (column~6), radio power at 5~GHz (column~7), 
extended radio power at 1.4~GHz 
as measured from FIRST images (column~8), R$^*$ parameter (column~9).

We discuss in detail the properties of the selected objects.

\begin{table*}
\begin{center}
\begin{tabular}{l r r r r r r r r}
Name & Position (J2000) & $S_{1.4~GHz}$ & $S_{5~GHz}$ & $S_{8.4~GHz}$ & z &Log P$_{5~GHz}$ 
& Log P$^{ext}_{1.4~GHz}$ & {\it R$^*$} \\
\ & \ & (mJy) &  (mJy) & (mJy) & \ & (W Hz$^{-1}$) & (W Hz$^{-1}$) & \ \\ 
\hline
GBJ141343+433959 & 14 13 43.717 +43 39 45.01 & 48 & 39   & 27  & 0.089 & 24.13 & 22.56 & 50 \\
GBJ151838+404532 & 15 18 38.903 +40 45 00.21 & 44 & 44   & 27  & 0.065 & 23.90 & 22.35 & 31 \\ 
GBJ171914+485839 & 17 19 14.491 +48 58 49.44 & 145 & 164 & 206 & 0.024 & 23.61 & 22.51 & 31 \\

\hline
\end{tabular}
\end{center}
 
\caption{Objects selected for the EVN observation}

\end{table*}

\subsection{Properties of the selected sources}

{\bf GBJ141343+433959} This object is classified from the literature 
as Broad Line Radio galaxy  (BLRG). It is the cD galaxy of 
the Abell 1885 cluster. The optical spectrum published in 
Crawford et al. (1995) and in Laurent-Muehleisen (1998) shows 
relatively strong emission lines ([OII], H$\beta$, [OI], 
H$\alpha$+[NII] and [SII]). GBJ141343+433959 is also a bright
X-ray source detected in the Bright ROSAT All Sky Survey (RASS,
Vogues et al. 1999). Its unabsorbed X-ray flux is 4.9$\times$10$^{-12}$ 
erg s$^{-1}$ cm$^{-2}$) in the 0.2--2.4 keV energy band corresponding
to an X-ray luminosity of 1.7$\times$10$^{44}$ erg s$^{-1}$. 
This high X-ray luminosity is probably due to the hot intercluster
plasma. The object belongs also to the REX catalogue (Radio Emitting
X-ray sources, Caccianiga et al. 1999).

\vspace*{0.5 cm}
\noindent
{\bf GBJ151838+404532} This object is classified from the literature 
as type~1 Seyfert galaxy. Its optical spectrum (Laurent-Muehleisen et al. 1998)
shows emission lines (H$\beta$, [OIII], H$\alpha$). As with the previous
source, GBJ151838+404532 was also detected in the Bright RASS catalogue
with  an unabsorbed flux of 8.4$\times$10$^{-12}$ 
erg s$^{-1}$ cm$^{-2}$ and an X-ray luminosity of
1.6$\times$10$^{43}$ erg s$^{-1}$. 

\vspace*{0.5 cm}
\noindent
{\bf GBJ171914+485839} This object belongs to an interacting 
system (Arp 102A and Arp 102B) where Arp 102B is the object
that shows AGN activity. The optical spectrum (e.g. Stauffer et
al. 1983) shows double-peaked broad emission lines (H$\beta$ and
H$\alpha$) superimposed on  narrow components. It is an
X-ray source detected in the bright RASS catalogue with a
unabsorbed flux of 9.2$\times$10$^{-12}$ erg s$^{-1}$ cm$^{-2}$
and an X-ray luminosity of 2.3$\times$10$^{42}$ erg s$^{-1}$.

In the radio band, GBJ171914+485839 has been detected as a compact
source at VLBI resolution (at 5~GHz) by Biermann et al. (1981) 
with a total flux of 96~mJy and an upper limit on the size of 0.4 pc.
At 4.9~GHz there is also a VLA observation in A-configuration
(Puschell et al. 1986) that shows a total flux of 90~mJy mainly 
concentrated in a unresolved core. About 30\% of this flux, however,
is distributed in a faint jet-like structure. Based on other
single-dish observations, Puschell et al. (1986) concluded that any
extended emission, if present, had to be much weaker than the
typical extended emission in ``normal'' radio galaxies.

\section{EVN observations}

\begin{figure}
\centerline{GBJ141343+433959}
\psfig{figure=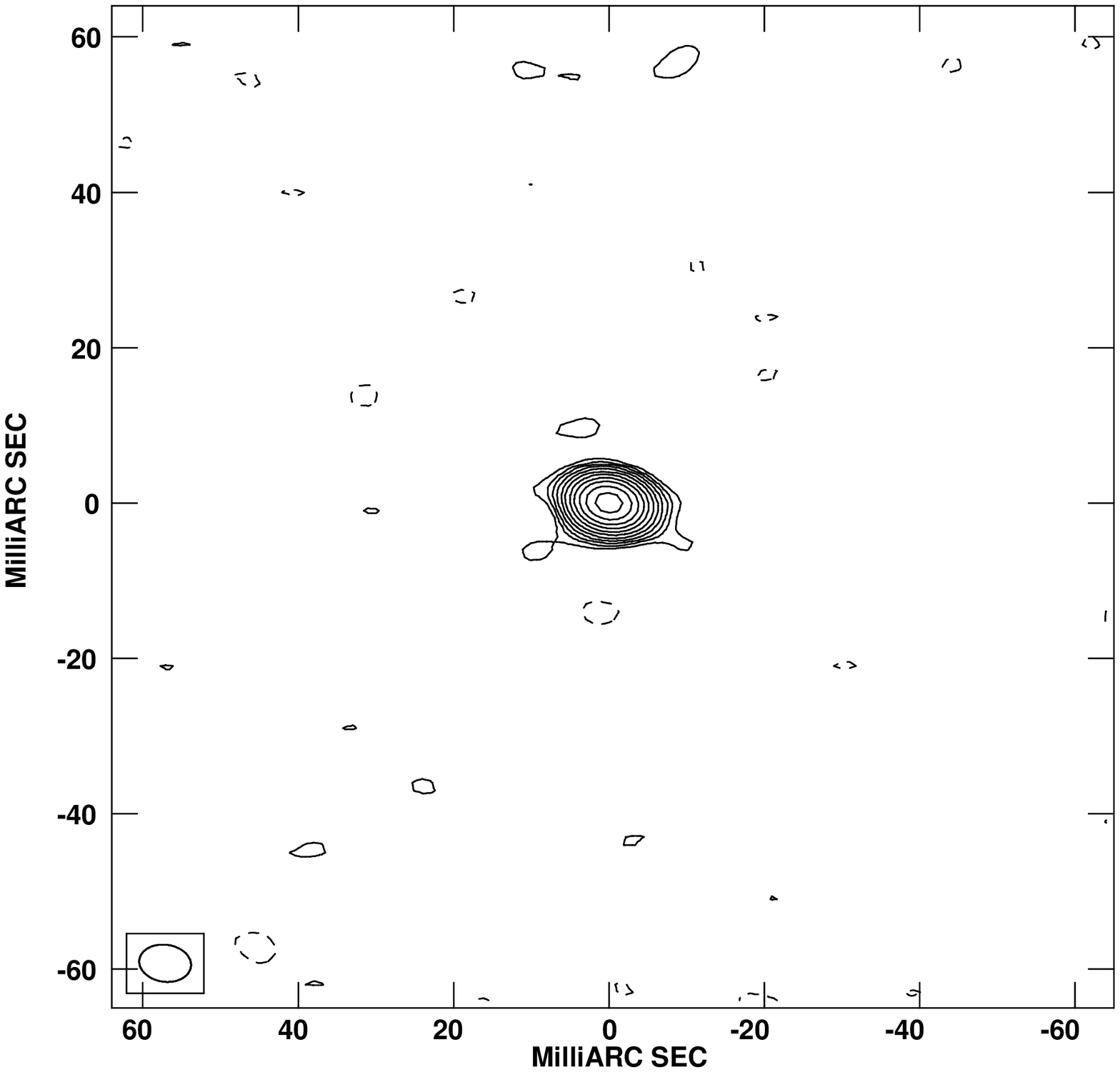,height=7cm,width=7cm}
\vspace*{0.5cm}
\centerline{GBJ151838+404532}
\psfig{figure=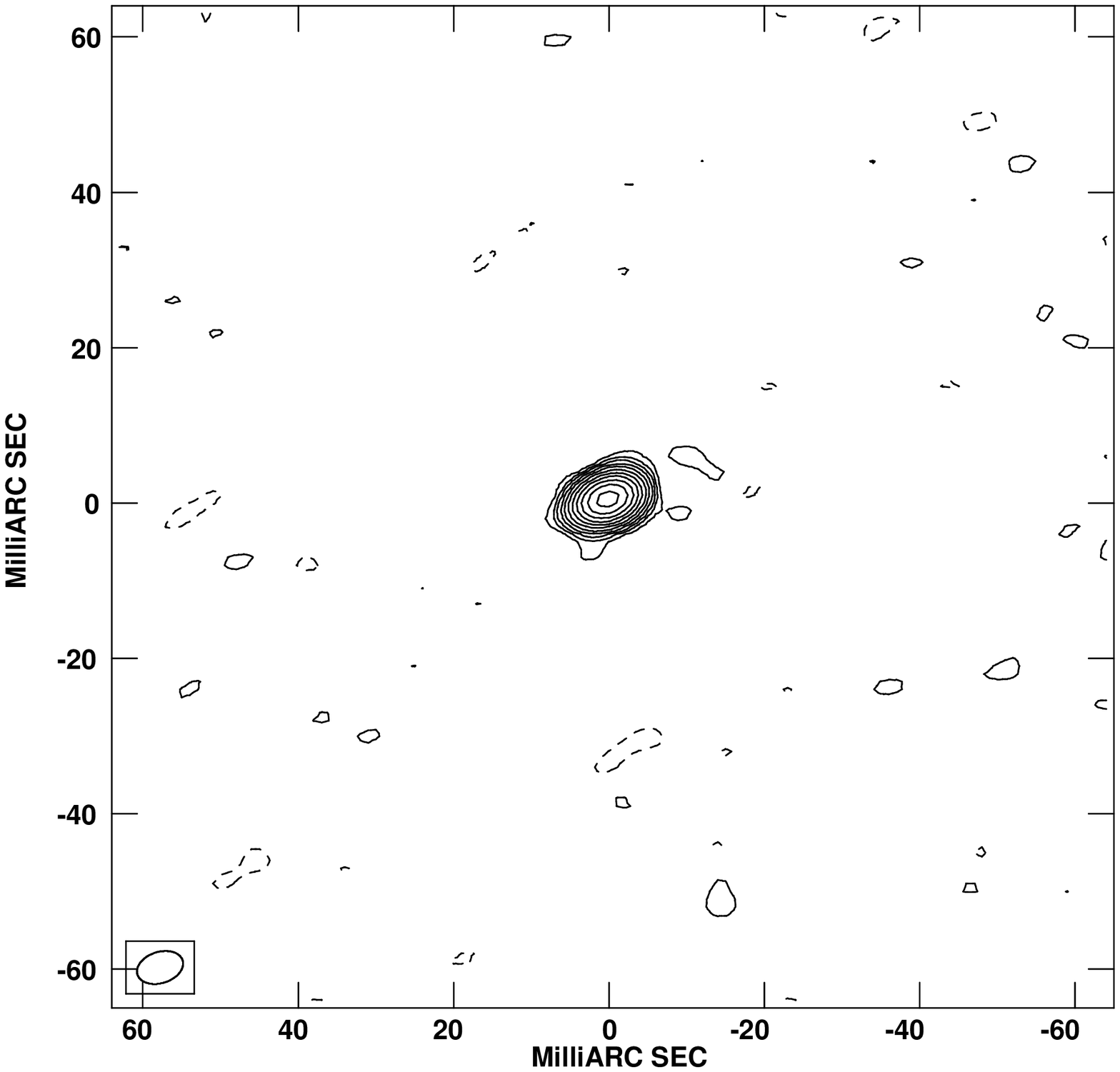,height=7cm,width=7cm}
\vspace*{0.5cm}
\centerline{GBJ171914+485839}
\psfig{figure=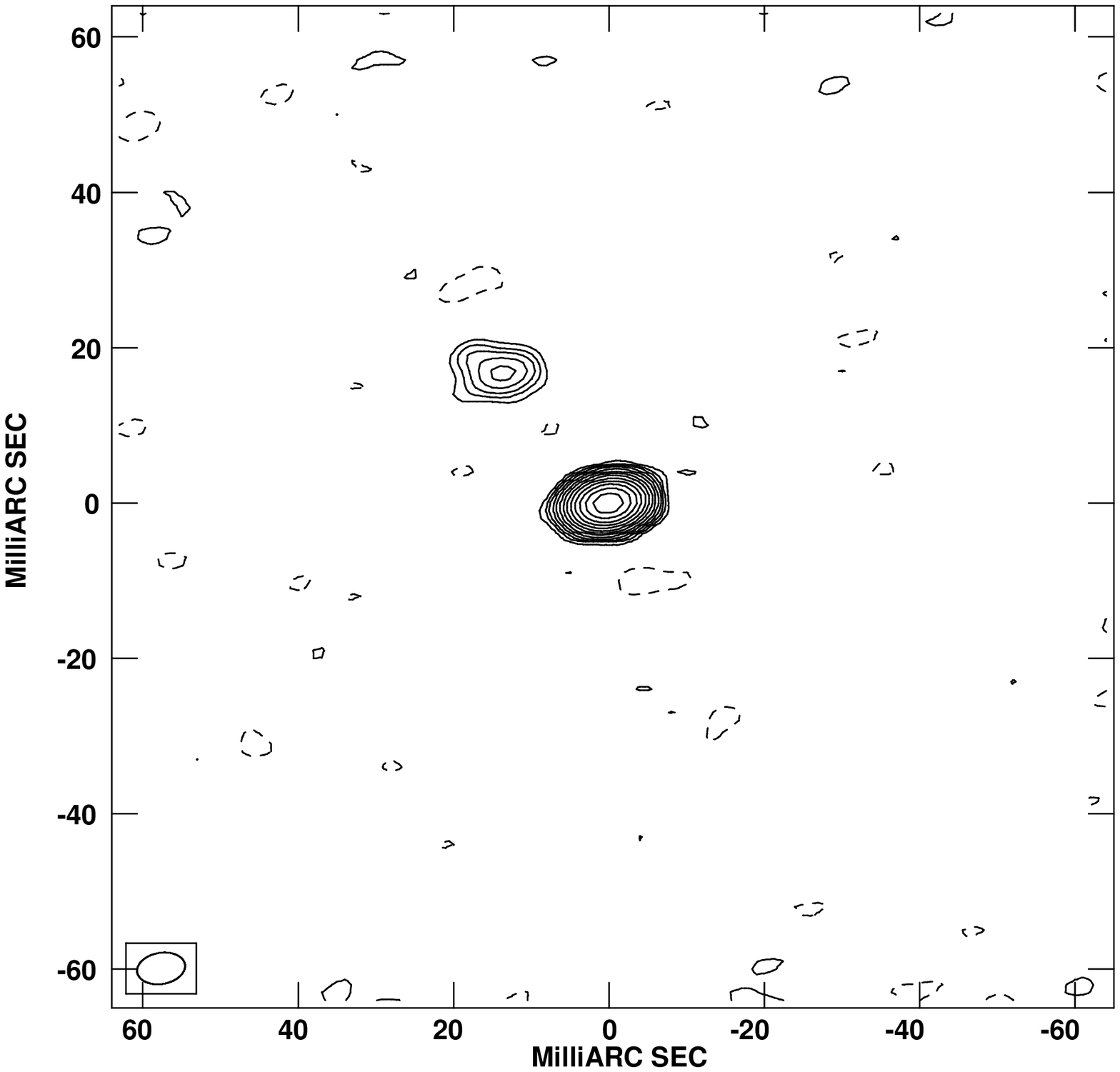,height=7cm,width=7cm}
\caption{EVN maps at 5~GHz of the CLASS RQ/RI AGNs.
The contour levels are: --2.5, 2.5, 4, 5, 5.6, 8, 11.3, 
16, 23, 32, 45, 64, 90, 128, 181 and they are 
counted in units of rms value (see Table~2)}
\label{maps}
\end{figure}

The three objects
have been observed at 5~GHz with the  European VLBI Network (EVN) on 
the 4$^{th}$ 
of June 2000 in VLBA recording mode. 
Seven antennas were used for these observations (Effelsberg, Westerbork,
Jodrell Bank, Medicina, Noto, Onsala, Torun). 
The total exposure time for each object was about 2 hours.
The data have been reduced at JIVE using the NRAO AIPS package.

\begin{table*}
\begin{center}
\begin{tabular}{l r r r r r r}
Name & Peak flux density & Total flux density & Beam size & rms & PA & T$_B$ \\
\ & (mJy beam$^{-1}$) &  (mJy) & (m.a.s.) & (mJy beam$^{-1}$) & (Deg) & (K) \\ 
\hline
GBJ141343+433959     & 32.0 & 33.4 & 6.7 x 4.8 & 0.29 &  +82.2 & $>$2.9$\times$10$^8$ \\
GBJ151838+404532     & 24.7 & 27.6 & 6.1 x 4.0 & 0.24 & --72.4 & $>$2.9$\times$10$^8$ \\ 
GBJ171914+485839 (A) & 54.9 & 55.6 & 6.2 x 4.0 & 0.23 & --80.4 & $>$6.4$\times$10$^8$ \\
GBJ171914+485839 (B) &  2.9 &  4.6 & 6.2 x 4.0 & 0.23 & --80.4 & $\sim$4$\times$10$^6$   \\

\hline
\end{tabular}
\end{center}
 
\caption{Results from EVN observations}
 
\end{table*}

The resulting maps are presented in Figure~\ref{maps} 
while in Table~2 we report, respectively:
The source's name (column~1), the peak flux density (column~2),
the total flux density (column~3), the beam size (column~4),
the rms (column~5), the position angle (column~6), 
the computed brightness temperature (or lower limit, column~7).
  
The noise for the 3 maps ranges from 0.23 - 0.29 mJy/beam 
which is relatively low and close to the theoretical value. 
The three objects appear compact without any evidence of
extended structure. Only in one case (GBJ171914+485839)
a second component is detected at a distance of 21.9 m.a.s. 
away from  the main component. The corresponding 
linear projected distance at the redshift of the source 
is about 15~pc.
In 1981 this
galaxy was observed with VLBI at 5~GHz, including Effelsberg, 
Westerbork and Green Bank (USA) antennas (Biermann et al. 1981). 
The object appeared unresolved in these observations and
the second component was not seen. 
However, even though the resolution
of the 1981 VLBI observation was good enough to separate 
the two components, the rms was probably too large 
($\sim$9 - 14 mJy) to clearly detect the faintest peak.  

Assuming an apparent  superluminal velocity for this 
component, a second epoch observation taken two
years after our observations should show an increment
in the relative distance between the two components of
more than 0.8~m.a.s.. This increment should be clearly 
detectable with observations at the same resolution 
(beam size of $\sim$ 4-7 m.a.s.). 

\subsection{Brightness temperature limits}
The simple fact that the 3 sources are detected 
and unresolved at the EVN resolution implies quite
a high lower limit on the brightness temperature. 
At the frequency of 5~GHz the brightness temperature
is given by:

\begin{equation}
T_B = 7.2 \times 10^7 \frac{S_{5~GHz}}{\Theta^2} K 
\end{equation}

where $S_{5 GHz}$ and $\Theta$ represent the observed flux
density (in mJy) and the source size (in m.a.s.) respectively.
Since all the detected components are unresolved (except for
the second weak component in GBJ171914+485839) their size is expected
to be lower than the beam size. A reasonable upper limit 
to the source size is  half of the beamsize. By using this
upper limit and the observed peak flux density 
we derive a lower limit for  T$_B$ 
(see Table~2). In general, for the 3 strongest components
in the 3 objects the lower limit on T$_B$ is larger than 
10$^8$~K. 

High values of brightness temperature are usually
interpreted as  strong indication for a non-thermal origin
for the observed radio emission. Recently Smith, Lonsdale \& Lonsdale 
(1998) investigated the possibility that high-T$_B$ and compact nuclei
observed in a sample of luminous infrared galaxies (LIG) are produced
by luminous radio supernovae generated by intense starburst activity. 
They found that in some cases the model can explain
what is observed at VLBI resolution although the supernovae luminosity 
has to be extremely high and, in some cases, more than an order 
of magnitude higher than the supernovae discovered so far. 
It must be noted, however, that the lower limits on T$_B$ derived for
the objects studied by Smith, Lonsdale \& Lonsdale (1998)  
are at least one order of magnitude below the 
ones inferred for the 3 sources presented in this paper. Moreover,
the well studied  Arp~220, which is considered the prototype luminous infrared 
galaxy, when observed at VLBI (with a resolution similar to that achieved
in our observations) shows a nucleus which is resolved 
into multiple components (Smith et al. 1998) something that is not
observed in the 3 sources presented here. Finally, there is
also a difference in radio power  
between the LIGs and the 3 sources presented here: the VLBI 
radio power (at 1.6~GHz) measured for the LIGs in the 
Smith, Lonsdale \& Lonsdale  sample is below 
10$^{23}$ W Hz$^{-1}$ (except for MKN 231) while the radio power
of the 3 CLASS sources measured at 5~GHz is above 10$^{23}$ W Hz$^{-1}$. 

We thus consider it unlikely that a model based on luminous radio supernovae
can explain the high T$_B$ observed in the 3 sources presented in 
this paper and we favour the hypothesis that we are observing a non-thermal 
emission from the AGN nucleus.

\subsection{Variability}

By comparing the total flux densities at 5~GHz 
in Table~1 and Table~2 we see that 
in GBJ151838+404532 and in GBJ171914+485839 the
total fluxes measured with EVN are significantly 
below the fluxes  in the GB6 catalog. 
This could be an indication either of variability
or that some flux is missed in the EVN observations.
However, in the case of  GBJ151838+404532, a second 
(unrelated) source  about 2 arcmin northward
is probably contributing to the GB6 flux, since
the large beamsize of GB ($\sim$3.5 arcmin) cannot
resolve the two sources. Furthermore, Laurent-Muehleisen et al. (1997) 
give a flux density of 29~mJy at 5~GHz for this source
measured with the VLA (A-configuration). This value is in 
good agreement with the one measured in the EVN
observation (27.6 mJy). 

In the case of  GBJ171914+485839, clear evidence of 
variability is given by the 
different measurements at 5~GHz taken at different
epochs (see Table~3), in particular by comparing 
the VLBI observations of Biermann et al. (1981)
to ours. The observed variability is 
consistent with the hypothesis that this source
is relativistically beamed.

\begin{table*}
\begin{center}
\begin{tabular}{l l rl}
Date & Telescope & S$_{5 GHz}$ & Ref. \\
\ &  &  (mJy) &   \\ 
\hline

1981	   & VLBI            & 96  & Biermann et al. (1981) \\
1981/03/10 & VLA (A config.) & 90 & Puschell et al. (1986) \\
1994/05/07 & VLA (B config.) & 139 & Laurent-Muehleisen et al. (1997) \\
2000/05/04 & EVN             & $^a$60  & This paper \\

\hline
\end{tabular}
\end{center}
 
\caption{Measurements of flux density at 5~GHz for GBJ171914+485839 at 
different epochs}

$^a$ this is the sum of the flux densities of the two components (A and B).
\end{table*}

\section{Radio-quiet or weak Radio-loud AGNs?}

As previously discussed, the 3 sources presented in this
paper were classified as RQ AGN on the basis of the observed
extended radio power at 1.4~GHz. However, the distinction between
RL and RQ AGN based on the radio-power is not universally 
accepted. In a recent paper Ho \& Peng (2001) 
suggested that a large fraction  of the Seyfert galaxies, usually 
considered as the local counterparts of the RQ QSOs, are 
actually RL objects according to the Kellerman definition 
(nuclear R$>$ 10) if the proper contribution from the nucleus
is considered. If confirmed, this result would imply that 
the RL class of AGN extends to orders of magnitude below the
usual dividing power of 10$^{25}$ W Hz$^{-1}$ and has a 
large overlap with the range of radio power shown by 
``real'' RQ AGNs (i.e. objects with a nuclear R below 10). 

In this case, the objects selected in the CLASS survey could 
be RL or RQ objects depending on the actual value of nuclear R parameter.
As already discussed, the computed R$^*$ is just an
upper limit and cannot be used to distinguish between RQ and RL
objects. 

Ho \& Peng (2001) also found two independent correlations between
the nuclear radio power (P$_{5 GHz}$) and line luminosity (L(H$\beta$)) 
for RQ and RL objects.
In principle, the location of an object in the P$_{5 GHz}$/L(H$\beta$)
plane could be used to distinguish between RL and RQ objects. 
The line luminosities (H$\beta$) and the radio powers at 5~GHz
of the 3 objects presented here are 
$\sim$10$^{40}$ - 10$^{41}$ erg s$^{-1}$ and  
$\sim$10$^{23}$ - 10$^{24}$ W Hz$^{-1}$ respectively. The position 
of these sources in the P$_{5~GHz}$/L(H$\beta$) plot is 
above the two best fits obtained by Ho \& Peng (2001) for RL and RQ AGN
respectively.
This is expected if the observed radio power is relativistically 
beamed. Depending on the value of the beaming parameter, 
the 3 objects can move down in the P$_{5~GHz}$/L(H$\beta$) plane   
to the RL correlation line, if a beaming
factor of 10 - 100 is considered, or to the RQ correlation line, 
by considering a  beaming factor of 100 - 1000. 
In the case of BL Lac objects, which are typical beamed 
RL AGN, the  amplification 
factor is estimated to be around 100 (e.g.  Urry \& Padovani 1995) 
but with a large spread around this value depending
on the plasma bulk velocity ($\Gamma$), the viewing angle ($\theta$)
and the fraction of luminosity emitted along the jet ($f$). 
Thus, both the RL and the RQ classifications are possible 
although the hypothesis that
the 3 objects presented here belong to the faint tail of the RL
population is probably the most reasonable since it does not require
large amplification factors.

An estimate of the beaming factor, derived from future detections
of SL motions in the selected objects, could be used to find the 
intrinsic radio power and, thus, determine the location
of the objects in the P$_{5~GHz}$/L(H$\beta$) diagram. 
This can give an indication of whether the objects belong to 
the RQ class of AGN or to the faint tail of the RL objects.

\section{Summary and conclusion}

Starting from a radio survey of flat-spectrum radio sources
we have selected 3 objects that show weak extended radio components whose 
powers are in the range of a typical RQ AGN ($<5\times$10$^{22}$ W Hz$^{-1}$).
In addition, these objects are at low-redshift
and their optical spectra show broad emission
lines. According to some authors (e.g. Terlevich et al. 1992)
the radio emission could be produced by thermal mechanisms,
like a high star formation rate in a dense environment. 
The EVN observations presented here, however, show in 
each target a bright and unresolved radio core whose 
brightness temperature ($>$ 10$^8$ K) is too high to
be due to starburst emission. Hence, non-thermal emission
(maybe due to the presence of a jet-like structure like
in more powerful RL AGN), is  probably causing the observed 
radio emission. 

The main reason we undertook this study was to investigate
the hypothesis that low-power AGNs, usually classified as RQ, 
can produce relativistic jets. To this
end we have focused our attention on the population of sources
for which direct evidence of relativistic outflows 
ought to be obtained most easily, namely the low-luminosity
flat-spectrum AGN. These objects could be RQ AGN
whose relativistic jets are viewed close to end-on.
Studying  the proper motions of radio components in a large
enough sample of such sources should uncover apparent 
superluminal motions if RQ AGN produce relativistic 
jets. 

Our initial results are promising: all 3 sources we observed
contain high-brightness temperature cores which can be 
monitored. One object (GBJ171914+485839) is particularly 
interesting because it is resolved into two components: 
superluminal motion in this source would be apparent within  
two years. 

The discovery that some low-power AGNs have relativistic jets
would either imply that RQ AGNs are intrinsically similar to 
RL objects or it may support the hypothesis, recently presented 
by Ho \& Peng (2001), that the RL class of AGN extends down to
very low radio powers and it overlaps significantly with 
the RQ AGN class. Both hypothesis have interesting implications
for the RL/RQ dichotomy debate.

\section*{Acknowledgments}
We thank  the staff at the JIVE institute, and in particular 
Denise Gabuzda, for their friendly support during all
the phases of EVN observation (proposal preparation, scheduling and 
data analysis). 
We acknowledge support from the European Commission under the IHP
Programme (ARI) contract No. HPRI-CT-1999-00045.
This research was supported in part by the European Commission
Training and Mobility of Researchers (TMR) program, research 
network contract ERBFMRX-CT96-0034 ``CERES''.

{}

\end{document}